\documentclass[aps,prb,twocolumn,superscriptaddress,floatfix,showpacs,10pt,letterpaper]{revtex4-1}

\pdfoutput=1

\usepackage{graphicx}
\usepackage{dcolumn}
\usepackage{bm}
\usepackage{amsfonts}
\usepackage{amsmath}
\usepackage{multirow}

\begin{document}

\begin{titlepage}

\title{2D symmetry protected topological orders and their protected gapless edge excitations}

\author{Xie Chen}
\affiliation{Department of Physics, Massachusetts Institute of
Technology, Cambridge, Massachusetts 02139, USA}

\author{Zheng-Xin Liu}
\affiliation{Department of Physics,
Massachusetts Institute of Technology, Cambridge, Massachusetts
02139, USA}
\affiliation{Institute for Advanced Study, Tsinghua University,
Beijing, 100084, P. R. China}

\author{Xiao-Gang Wen}
\affiliation{Department of Physics, Massachusetts Institute of
Technology, Cambridge, Massachusetts 02139, USA}
\affiliation{Institute for Advanced Study, Tsinghua University,
Beijing, 100084, P. R. China}

\begin{abstract}
Topological insulators in free fermion systems have been well characterized and
classified. However, it is not clear in strongly interacting boson or fermion
systems what symmetry protected topological orders exist. In this paper, we
present a model in a 2D interacting spin system with nontrivial on-site $Z_2$
symmetry protected topological order. The order is nontrivial because we can
prove that the 1D system on the boundary must be gapless if the symmetry is not
broken, which generalizes the gaplessness of Wess-Zumino-Witten model for Lie
symmetry groups to any discrete symmetry groups. The construction of this model
is related to a nontrivial 3-cocycle of the $Z_2$ group and can be generalized
to any symmetry group. It potentially leads to a complete classification of
symmetry protected topological orders in interacting boson and fermion systems
of any dimension.  Specifically, this exactly solvable model has a unique
gapped ground state on any closed manifold and gapless excitations on the
boundary if $Z_2$ symmetry is not broken. We prove the latter by developing the
tool of matrix product unitary operator to study the nonlocal symmetry
transformation on the boundary and revealing the nontrivial 3-cocycle structure
of this transformation.  Similar ideas are used to construct a 2D fermionic
model with on-site $Z_2$ symmetry protected topological order.

\end{abstract}


\maketitle

\vspace{2mm}

\end{titlepage}

\section{Introduction}

Topological phases of matter are gapped quantum systems containing nontrivial
orders which are not due to spontaneous symmetry breaking in the ground states.
While topologically ordered systems all have exponentially decaying correlation
and appear quite simple from a classical point of view, various exotic quantum
features have been discovered which reveal the surprisingly rich structure of
topologically ordered systems. For example, some systems have ground state
degeneracy which depends on the topology of the closed manifold the system is
on;\cite{W8987,WN9077} some have protected gapless edge excitations if the
system has a boundary;\cite{W9038,W9125} some have nontrivial entanglement
structure in the ground state;\cite{KP0604,LW0605} and some have bulk
excitations with nontrivial statistics.\cite{ASW8422,K0602,LW0510} How to
obtain a clear picture of topological phases among such a variety of phenomena?
First we find that topological phases can be divided into two general classes
according to its level of stability under perturbations. 

The first class has `intrinsic' topological order.\cite{W8987} Systems in this
class must go through a phase transition to a trivial phase no matter what kind
of local perturbation is added. Or using the local unitary equivalence between
ground states we find that this class of systems have ground states which
cannot be mapped to a product state under ANY local unitary transformation as
defined in \Ref{CGW1038}. We say that this kind of states have long range
entanglement.  Example systems in this class include quantum Hall(integer or
fractional),\cite{TSG8259,L8395} $p+ip$ superconductor,\cite{RG0067,GR0709}
string-net models,\cite{LW0510} $Z_2$ spin liquid,\cite{RS9173,W9164,MS0181}
and chiral spin liquid.\cite{KL8795,WWZ8913} It has been discovered that
systems with `intrinsic' topological order usually have topology dependent
ground state degeneracy,\cite{W8987,WN9077} nontrivial topological entanglement
entropy\cite{KP0604,LW0605} and fractional statistics of bulk
excitation.\cite{ASW8422,K0602,LW0510} In the following discussion we will use
the term `topological order' to specifically refer to this class of systems.

The second class has `symmetry protected' topological order. This kind of
system has certain symmetry and its non-degenerate ground state does not break
any of the symmetries. If arbitrary perturbations are allowed, systems in this
class all belong to the same phase as a trivial state. Its ground state can be
mapped to a product state with local unitary transformations and hence are
short range entangled(SRE). However, if only symmetric perturbations are
allowed, systems in this class are in different phases from the trivial phase.
Therefore, we say that the topological order in this class is symmetry
protected. We will call these phases `symmetry protected topological' (SPT)
phases. Example systems in this class include Haldane phase in one dimensional
spin chain\cite{H8364} and topological
insulators.\cite{KM0501,BZ0602,KM0502,MB0706,FKM0703,QHZ0824} Systems with SPT
order have non-degenerate ground states on closed manifold and usually have
nontrivial edge degrees of freedom if the system has a
boundary.\cite{K9037,KM0501,BZ0602,KM0502,MB0706,FKM0703,QHZ0824}

Many efforts have been made to obtain a more complete understanding of
topological and symmetry protected topological orders. In particular,
topological and SPT orders have been completely classified in one-dimensional
spin systems.\cite{CGW1107,SPC1032} It was found that one-dimensional spin
systems cannot have nontrivial topological order but different SPT orders exist
for systems with certain symmetry. Similarly, a classification of fermion
systems (interacting) in one dimension is also
possible.\cite{LK1103,TPB1102,CGW1123} The picture changes dramatically in
higher dimensions. First of all, nontrivial topological order does exist in two
or higher dimensions. A lot has been learned about possible topological
orders\cite{WZ9290,K0302,LW0510,CGW1038,GWW1017} although a complete
understanding is still missing. In this paper, we are going to focus only on
the SPT phases. Most SPT phases in two and higher dimensions have been
identified in free fermion systems due to the simplicity and versatility of the
formalism. A classification of possible SPT phases in non-interacting fermion
systems has been obtained.\cite{SRF0825,RSF1010,K0922} The major open question
about SPT phases is in general which of these phases remain and what new SPT
phases are possible when the system is strongly interacting. In boson systems,
even less is known as non-interacting bosons are necessarily topologically
trivial. \footnote{Recently, there are several proposals for `fractional
topological insulators',\cite{LS0903,SBM1076,MQK1009,LBK1154} which incorporate
interaction effect into topological insulators and find topologically ordered
phases. However, these phases all have intrinsic topological order and does not
belong to SPT phases discussed here.}

In this paper, we present a generic picture for understanding SPT phases in
interacting systems through the explicit construction of a simple example.
Instead of starting from free fermions, we take a different approach and
generalize our understanding of one dimensional interacting SPT phases to
construct a two dimensional spin model with on-site $Z_2$ symmetry protected
topological order. We call this model the CZX model for reason that will become
clear later. On a closed surface the CZX model looks simple. Its Hamiltonian is
composed of commuting projectors. Its symmetric gapped ground state is a
product of local loops and hence short range entangled. However, the model
becomes highly nontrivial if it has a boundary. The boundary must have gapless
excitation as long as symmetry is not broken, a signature of nontrivial SPT
order. We prove this fact by relating effective symmetry transformation on the
boundary with a nontrivial 3-cocycle of the $Z_2$ group. 

The construction of the CZX model signifies the close relation between SPT
phases and nontrivial cocycles of the symmetry group. This idea is not limited
to two dimensional systems. In another paper,\cite{CGLW1172} we generalize the
formalism and construct nontrivial SPT phases in any $d$ dimension with on-site
unitary and anti-unitary symmetries $G$ based on $(d+1)$-cocycles of $G$. We
expect that this construction gives a complete classification of
$d$-dimensional SPT phases.

The effective theory on the boundary can be seen as a generalization of the
Wess-Zumino-Witten (WZW) model.\cite{WZ7195,W8322} The WZW model describes
conformally invariant 1D systems with an internal symmetry of a compact Lie
group. The WZW model obtained by adding a topological term (the WZW term) to
the usual dynamical term in the Lagrangian of the nonlinear sigma model,  is
exactly solvable in semiclassical limit.  It explains the physics of 1D gapless
systems with a global Lie group symmetry. However, the construction of the
model depends crucially on the fact that the symmetry group is continuous and
does not apply to, for example, the $Z_2$ group. Our proof of the gapless-ness
of the 1D effective theory on the boundary of the CZX model hence generalizes
the understanding of the WZW model to discrete groups. Our method based on the
nontrivial 3-cocycles applies to both continuous and discrete symmetry groups,
although it does not give the conformal field theory of the system directly.
Also our proof is non-perturbative, not relying on semiclassical approximation.
The connection between the CZX model and the WZW model is not particularly
clear in the formulation of this paper, as the WZW model is usually given in
the Lagrangian form. In another paper\cite{CGLW1172}, we reformulate our models
(including the CZX model and those for all other symmetries and in all
dimensions) in the Lagrangian language where the connection with the WZW model
would become obvious.

The paper is organized as follows: in section \ref{1-2D}, we review our
understanding of the entanglement structure of SPT phases in one dimension. In
generalizing such entanglement structure to higher dimension, we first present
a naive attempt which fails to produce interesting phases. Identifying the
missing element, we construct the CZX model in section \ref{CZX}. We give
explicitly the symmetry of the system, its Hamiltonian and its ground state. In
order to show the nontrivial-ness of this model, we study its effective
boundary theory in section \ref{CZX_bd}. We identify the effective degrees of
freedom, effective $Z_2$ symmetry and show that in simple cases the boundary
cannot be in a gapped symmetric phase. In order to prove this conclusion in
general, we use the tool of matrix product unitary operators(MPUO).
Introduction to the matrix product unitary operators formalism is given in
appendix \ref{MPUO} including its definition and some simple properties. In
section \ref{MPUO_H3}, we show how to represent the effective symmetry on the
boundary of the CZX model using MPUO. We find that the transformation rule
between the MPUO's is related to a nontrivial class of 3-cocycles in the third
cohomology group $\cH^3(Z_2,U(1))$ of $Z_2$.\footnote{the $(d+1)$-cohomology
group of $Z_2$ $\cH^{d+1}(G,U(1))$ is trivial for odd $d$ and is a $Z_2$ group
for even $d$} Using this relation, we prove that the boundary cannot have a
gapped symmetric ground state. This result applies in general to any MPUO
related to a nontrivial 3-cocycle in $\cH^3(G,U(1))$. Hence we conclude that
the CZX model is in a nontrivial SPT phase protected by on-site $Z_2$ symmetry.
Using similar ideas, we construct in section \ref{fermion} a fermion system
with on-site $Z_2$ symmetry whose boundary is also nontrivial. 

\section{From 1D SPT phases to 2D} \label{1-2D}

In this section we first review our understanding of the entanglement pattern
at the fixed point of 1D SPT phases which we then try to generalize to higher
dimensions. However, we are going to show that a straight forward
generalization fails to give nontrivial SPT order. We identify the missing
elements and prepare for the construction of nontrivial model in the next
section.

\begin{figure}[ht]
\begin{center}
\includegraphics[scale=0.4]{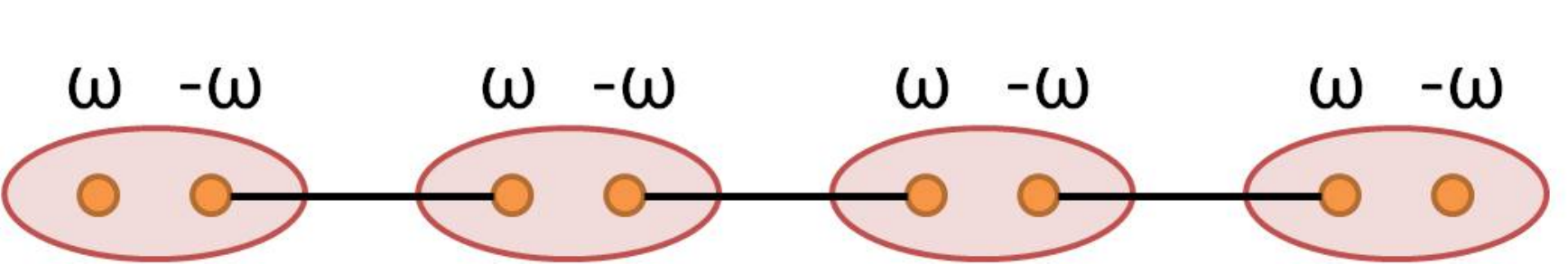}
\end{center}
\caption{Fixed point ground state of 1D SPT phase with on-site symmetry of group $G$. Each site contains two spins, which form projective representation of class $\omega$ and $-\omega$ respectively. Connected spins form a dimer which forms a one-dimensional representation of $G$. On a finite segment of the 1D chain, the boundary spins form projective representations of $G$.
}
\label{1D_fp}
\end{figure}

Each 1D SPT phase in systems with on-site symmetry $G$ can be well understood
from the entanglement pattern of its ground state at fixed point, as shown in
Fig.\ref{1D_fp}. At fixed point, each site contains two spins. On each site,
symmetry is represented linearly. But on each spin, symmetry only needs to be
represented projectively. (A simple example of projective representation is
given by $SO(3)$ symmetry on a spin $1/2$. For an introduction to projective
representations and the second cohomology group $\cH^2(G,U(1))$ see appendix
\ref{prorep}. More generally, group cohomology is introduced in appendix
\ref{Gcoh}.) If symmetry on the left spin belongs to the projective
representation of class $\omega$ in $\cH^2(G,U(1))$(for example spin $1/2$
under $SO(3)$), then on the right spin it belongs to $-\omega$(again spin $1/2$
under $SO(3)$) so that together they form a linear representation. The ground
state of the system is a product of dimers between spins on neighboring sites.
Each dimer is an entangled state of two spins which forms a one dimensional
representation of $G$. The ground state is hence a total singlet under the
symmetry. The nontrivial feature of the system shows up when we cut the chain
into a finite segment. There are free degrees of freedom at the ends of the
segment, each forming a projective representation of $G$. Two 1D systems belong
to the same SPT phase if their end degrees of freedom belong to the same class
of projective representation $\omega \in \cH^2(G,U(1))$.

\begin{figure}[ht]
\begin{center}
\includegraphics[scale=0.3]{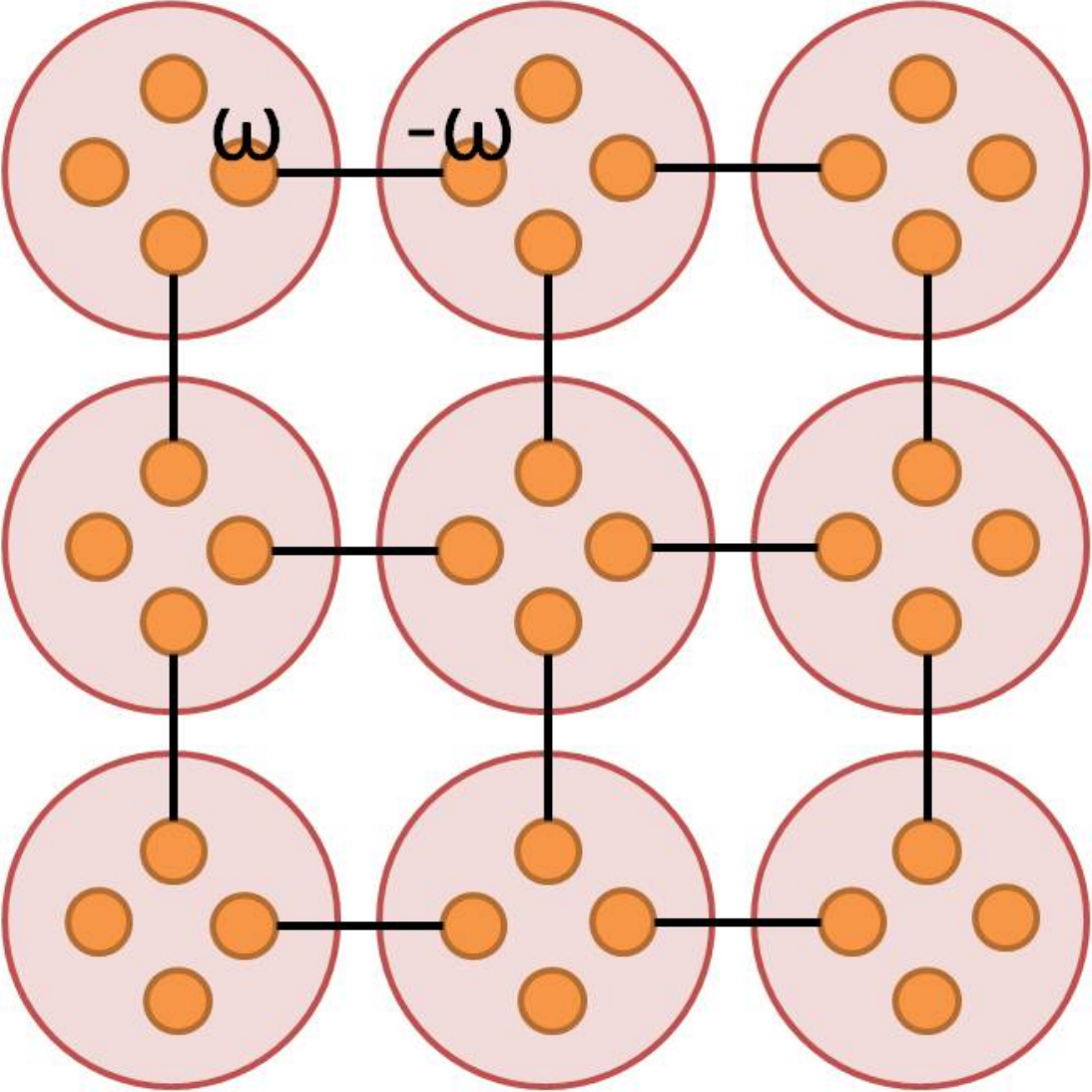}
\end{center}
\caption{A 2D `bond' state which is short range entangled and is symmetric under on-site symmetry of group $G$. Each site contains four spins, each forming a projective representation of $G$. Two spins connected by a bond form projective representations of class $\omega$ and $-\omega$ respectively. The `bond' represents an entangled state of the two spins which forms an one-dimensional representation of $G$. On a lattice with boundary, the boundary degrees of freedom are spins with projective representation $\omega$($-\omega$.)
}
\label{2D_bond}
\end{figure}

This simple picture can be generalized to two or higher dimension to give a `bond' state. Consider the 2D state in Fig. \ref{2D_bond}.

Every site contains four spins. Each spin forms a projective representation of on-site symmetry $G$, but the four spins on each site together form a linear representation of $G$. Two spins on neighboring sites which are connected by a bond forms projective representation $\omega$ and $-\omega$ respectively and the bond represents an entangled state between the two spins which forms a one dimensional representation of $G$. Similar to the 1D case, the total state is invariant under on-site symmetry $G$. The state is short range entangled and can be the gapped ground state of a simple Hamiltonian(sum of projections onto the entangled pairs). If the system is defined on a disk with boundary, there will be free degrees of freedom at each site on the boundary which form projective representations of $G$. 

It might seem that states with different projective representations at each site on the boundary correspond to different SPT phases, just like in the 1D case. However, this is not totally true. If translation symmetry is required, each boundary spin is well defined and the projective representation they form do label different phases. On the other hand, in the absence of translation symmetry, boundary spins can be combined and their projective representations can add together. As projective representations form an additive group (the second cohomology group $\cH^2(G,U(1))$ of $G$), combining boundary spins would change the projective representations from one class to another and in particular, to the trivial class. Therefore, without translation symmetry, all 2D states with a bond form as shown in Fig.\ref{2D_bond} belong to the same phase. 

On the other hand, SPT phases are known to exist in two and higher dimensions without the protection of translation symmetry, for example in topological insulators. The simple bond picture above therefore cannot account for their SPT order. In order to have nontrivial SPT order, we need to generalize the bond state in two ways: (1) the local entanglement structure is not bonds between two spins, but rather plaquettes among four spins on sites around a square. This alone is not enough to construct new SPT order. We also need (2) symmetry transformation on each site does not factorize into separate operations on each of the four spins. That is, the total linear symmetry operation on each site is not a tensor product of four projective representations as otherwise the state can be reduced to a bond state.

Following this line of thought, we construct the CZX model in section \ref{CZX}. The CZX model has an on-site $Z_2$ symmetry that does not factorize into projective representations and the symmetry protected topological order of the state is robust against disorder. The boundary effective degrees of freedom in CZX model has an effective $Z_2$ symmetry which cannot be written in an on-site form. Moreover, the boundary cannot be in a gapped symmetric state under the effective symmetry. In other words, the boundary must either break the $Z_2$ symmetry or have gapless excitations. This is different from the bond state discussed above(Fig.\ref{2D_bond}). In the bond state, the boundary degrees of freedom are the boundary spins with projective representations. The effective symmetry is still on-site. Several boundary spins can form a singlet if their projective representations add up to a linear representation. Therefore, in the bond state, the boundary can be in a gapped symmetric state under on-site symmetry simply by breaking translation symmetry. However, in the CZX model, this is not possible.

\section{CZX model}
\label{CZX}

In this section, we construct the CZX model explicitly which turns out to have nontrivial SPT order protected only by on-site $Z_2$ symmetry.

\begin{figure}[ht]
\begin{center}
\includegraphics[scale=0.5]{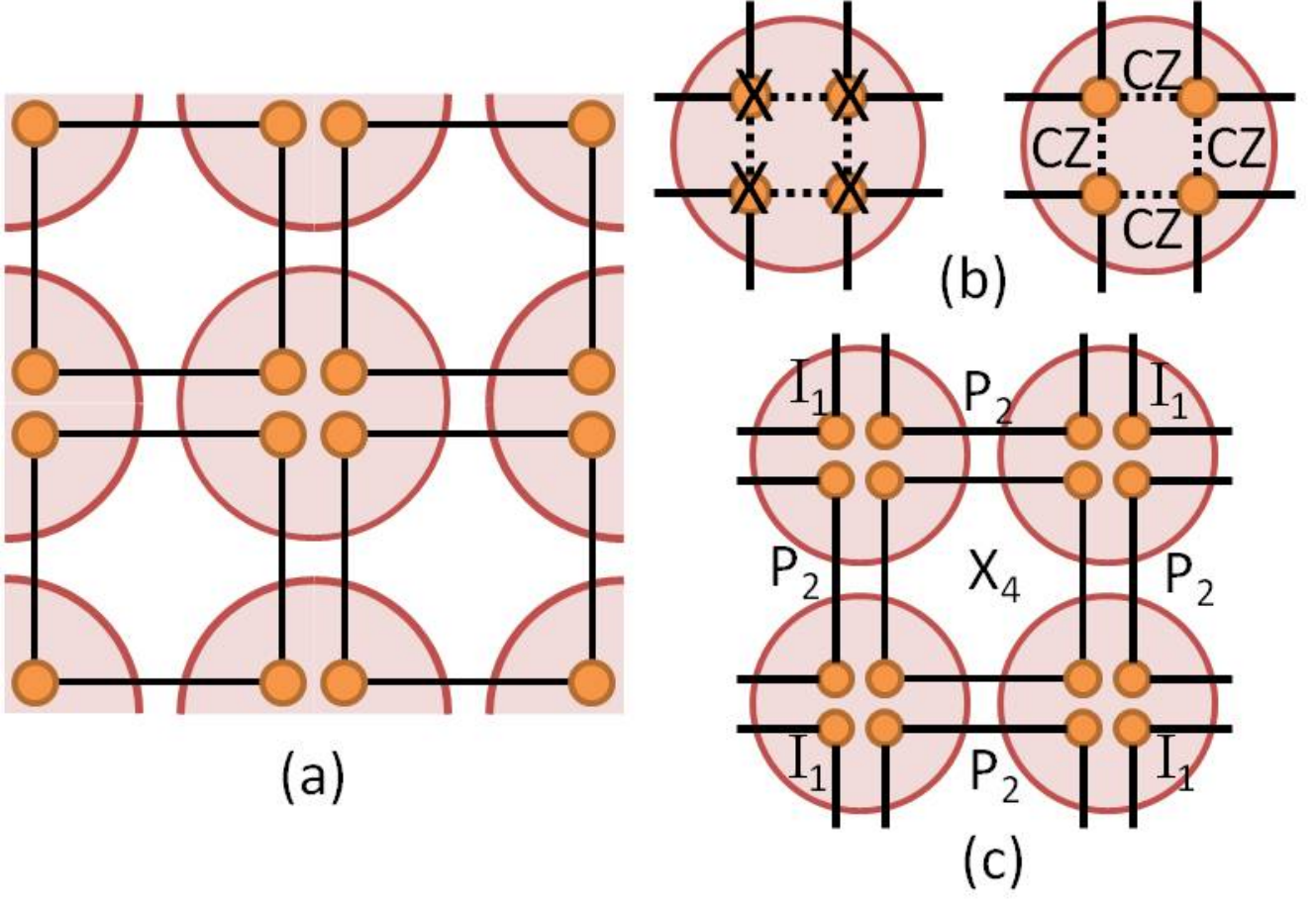}
\end{center}
\caption{CZX model (a) each site (circle) contains four spins (dots) and the spins in the same plaquette (square) are entangled. (b) on-site $Z_2$ symmetry is generated by $U_{CZX}=X_1 X_2 X_3 X_4CZ_{12}CZ_{23}CZ_{34}CZ_{41}$ (c) a local term in the Hamiltonian, which is a tensor product of one $X_4$ term and four $P_2$ terms as defined in the main text.
}
\label{CZX_model}
\end{figure}

Consider a square lattice with four two-level spins per site, as shown in Fig. \ref{CZX_model}(a) where sites are represented by circles and spins are represented by dots. We denote the two levels as $|0\>$ and $|1\>$. The system has an on-site $Z_2$ symmetry as given in Fig. \ref{CZX_model}(b). It is generated by
\be
U_{CZX}=U_X U_{CZ}
\ee 
where
\be
U_X=X_1\otimes X_2\otimes X_3\otimes X_4
\ee
$X_i$ is Pauli $X$ operator on the $i$th spin and 
\be
U_{CZ}=CZ_{12}CZ_{23}CZ_{34}CZ_{41}
\ee
where $CZ$ is the controlled-$Z$ operator on two spins defined as
\be
CZ=|00\>\<00|+|01\>\<01|+|10\>\<10|-|11\>\<11|
\ee
As defined, $CZ$ does nothing if at least one of the spins is in state $|0\>$ and it adds a minus sign if both spins are in state $|1\>$. Different $CZ$ operators overlap with each other. But because they commute, $U_{CZ}$ is well defined. Note that $U_{CZ}$ cannot be decomposed into separate operations on the four spins and the same is true for $U_{CZX}$. $U_{X}$ and $U_{CZ}$ both square to $I$ and they commute with each other. Therefore, $U_{CZX}$ generates a $Z_2$ group.

The Hamiltonian of the system is defined as a sum of local terms around each plaquette. Plaquettes are represented by squares in Fig. \ref{CZX_model}. $H=\sum H_{p_i}$, where the term around the $i$th plaquette $H_{p_i}$ acts not only on the four spins in the plaquette but also on the eight spins in the four neighboring half plaquettes as shown in Fig. \ref{CZX_model}(c)
\be
H_{p_i}=-X_4\otimes P_2^{u}\otimes P_2^{d}\otimes P_2^{l}\otimes P_2^{r}
\ee
where $X_4$ acts on the four spins in the middle plaquette as
\be
X_4=|0000\>\<1111|+|1111\>\<0000|
\ee
and $P_2$ acts on the two spins in every neighboring half plaquette as
\be
P_2=|00\>\<00|+|11\>\<11|
\ee
$P_2^{u}$, $P_2^{d}$, $P_2^{l}$, $P_2^{r}$ acts on the up, down, left and right neighboring half plaquettes respectively. For the remaining four spins at the corner, $H_{p_i}$ acts as identity on them. The $P_2$ factors ensure that each term in the Hamiltonian satisfies the on-site $Z_2$ symmetry defined before. 

All the local terms in the Hamiltonian commute with each other, therefore it is easy to solve for the ground state. If the system is defined on a closed surface, it has a unique ground state which is gapped. In the ground state, every four spins around a plaquette are entangled in the state 
\be
|\psi_{p_i}\>=|0000\>+|1111\>
\ee
and the total wavefunction is a product of all plaquette wavefunction. If we allow any local unitary transformation, it is easy to see that the ground state can be disentangled into a product state, just by disentangling each plaquette separately into individual spin states. Therefore, the ground state is short range entangled. However, no matter what local unitary transformations we apply to disentangle the plaquettes, they necessarily violate the on-site symmetry and in fact, the plaquettes cannot be disentangled if the $Z_2$ symmetry is preserved, due to the nontrivial SPT order of this model which we will show in the next sections.

It can be checked that this ground state is indeed invariant under the on-site $Z_2$ symmetry. Obviously this state is invariant under $U_X$ applied to every site. It is also invariant under $U_{CZ}$ applied to every site. To see this note that between every two neighboring plaquettes, $CZ$ is applied twice, at the two ends of the link along which they meet. Because the spins within each plaquette are perfectly correlated (they are all $|0\>$ or all $|1\>$), the effect of the two $CZ$'s cancel each other, leaving the total state invariant. 

Therefore, we have introduced a 2D model with on-site $Z_2$ symmetry whose ground state does not break the symmetry and is short-range entangled. In particular, this on-site symmetry is inseparable as discussed in the introduction and therefore cannot be characterized by projective representation as in the bond state. We can add small perturbation to the system which satisfies the symmetry and the system is going to remain gapped and the ground state short range entangled and symmetric. It seems that the system is quite trivial and boring. However, we are going to show that surprising things happen if the system has a boundary and because of these special features the system cannot be smoothly connected to a trivial phase even if translation symmetry is not required.

\section{CZX model boundary}
\label{CZX_bd}

\begin{figure}[ht]
\begin{center}
\includegraphics[scale=0.5]{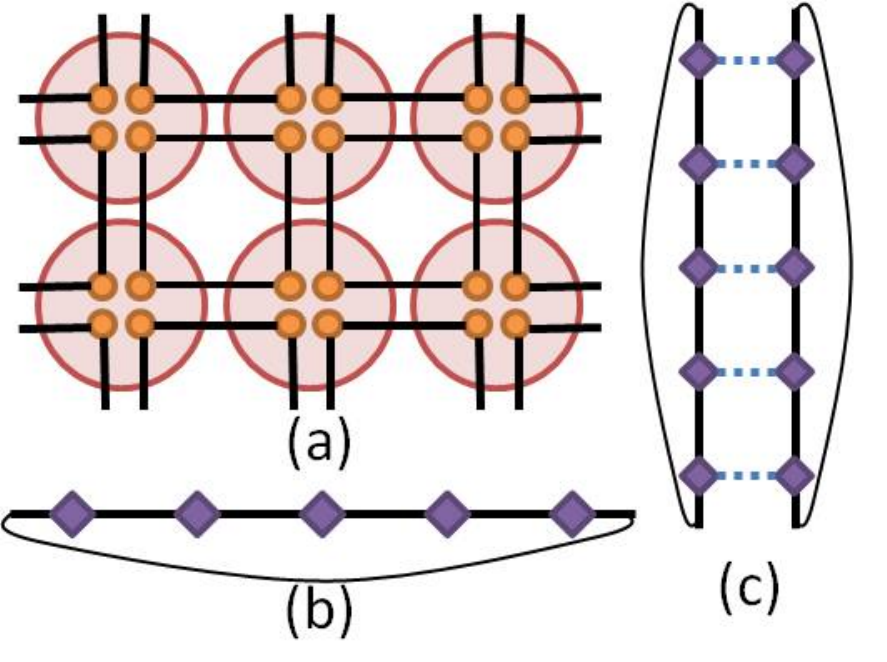}
\end{center}
\caption{(a)CZX model on a disk with boundary (b) boundary effective degrees of freedom form a 1D chain which cannot have a SRE symmetric state (c) two boundaries together can have a SRE symmetric state which is a product of entangled pairs between effective spins connected by a dashed line.
}
\label{CZX_boundary}
\end{figure}

The non-trivialness of this model shows up on the boundary. Suppose that we take a simply connected disk from the lattice, as shown in Fig.\ref{CZX_boundary}(a). 

The reduced density matrix of spins in this region is invariant under on-site symmetry in this region. The reduced density matrix is a tensor product of individual terms on each full plaquette, half plaquette and corner of plaquette respectively. On a full plaquette
\be
\rho_4=(|0000\>+|1111\>)(\<0000|+\<1111|)
\ee
On a half plaquette
\be
\rho_2=|00\>\<00|+|11\>\<11|
\ee
On a corner of a plaquette
\be
\rho_1=|0\>\<0|+|1\>\<1|
\ee
The state of spins on the plaquettes totally inside this region is completely fixed. But on the boundary there are free degrees of freedom. However, unlike in the bond state, only part of the total Hilbert space of the spins on the boundary is free. In particular, two spins in a half plaquette on the boundary are constrained to the two-dimensional subspace $|00\>\<00|+|11\>\<11|$ and form an effective spin degree of freedom if we map $|00\>$ to $|\t{0}\>$ and $|11\>$ to $|\t{1}\>$. 

In Fig. \ref{CZX_boundary}(b), we show the effective degrees of freedom on the boundary as diamonds on a line. Projecting the total symmetry operation on the disk to the space supporting reduced density matrix, we find that the effective symmetry operation on the boundary effective spins is $\t{U}_{CZX}=\prod_{i=1}^N \t{X}_i\prod_{i=1}^N \t{CZ}_{i,i+1}$, with Pauli $\t{X}$ on each effect spin and $\t{CZ}$ operation between neighboring effective spins. The boundary is periodic and $\t{CZ}_{N,N+1}$ acts on effective spin $N$ and $1$. This operator generates a $Z_2$ symmetry group.

This is a very special symmetry on a 1D system. First it is not an on-site symmetry. In fact, no matter how we locally group sites and take projections, the symmetry operations are not going to break down into an on-site form. Moreover, no matter what interactions we add to the boundary, as long as it preserves the symmetry, the boundary cannot have a gapped symmetric ground state. We can start by considering some simple cases. The simplest interaction term preserving this symmetry is $Z_iZ_{i+1}$. This is an Ising interaction term and its ground state breaks the $Z_2$ symmetry. In the transverse Ising model, the system goes to a symmetric phase if magnetic field in the $x$ direction is increased. However, $X_i$ breaks the $Z_2$ symmetry $\t{U}_{CZX}$ on the boundary and therefore cannot be added to the Hamiltonian. In fact, we are going to prove that the boundary cannot have SRE symmetric ground state (actually a more generalized version of it) in the next section. This is one special property that differs the CZX model from the bond state in Fig.\ref{2D_bond}. In the bond state, the symmetry operations on the boundary are just projective representations on each site. Without translational invariance, there can always be a SRE symmetric state with this symmetry. 

The special property on the boundary only shows up when there is an isolated single boundary. If we put two such boundaries together and allow interactions between them, everything is back to normal. As shown in Fig.\ref{CZX_boundary}(c), if we have two boundaries together, there is indeed a SRE symmetric state on the two boundaries. The state is a product of entangled pairs of effective spins connected by a dashed line. The entangled pair can be chosen as $|\t{0}\t{0}\>+|\t{1}\t{1}\>$. In contrast to the single boundary case, we can locally project the two effective spins connected by a dashed line to the subspace $|\t{0}\t{0}\>\<\t{0}\t{0}|+|\t{1}\t{1}\>\<\t{1}\t{1}|$ and on this subspace, the symmetry acts in an on-site fashion.

This result should be expected because if we have two pieces of sheet with boundary and glue them back into a surface without boundary, we should have the original SRE 2D state back. Indeed if we map the effective spins back to the original degrees of freedom $|\t{0}\>\to|00\>$ and $|\t{1}\> \to |11\>$, we see that the SRE state between two boundaries is just the a chain of plaquettes $|0000\>+|1111\>$ in the original state.

This model serves as an example of non-trivial SPT order in 2D SRE states that only needs to be protected by on-site symmetry. In order to prove the special property on the boundary of CZX model and have a more complete understanding of possible SPT orders in 2D SRE states with on-site symmetry, we are going to introduce a mathematical tool called Matrix Product Unitary Operator. We will show that 2D SPT phases are related to elements in $\mathcal{H}^3(G,U(1))$ which emerge in the transformation structure of the matrix product unitary operators. The definition of matrix product unitary operator and some basic properties are given in appendix \ref{MPUO}. The discussion in the next section is general, but we will work out the CZX example explicitly for illustration.

\section{Matrix Product Unitary Operators and its relation to 3 cocycle}
\label{MPUO_H3}

In this section, we discuss the matrix product unitary operator (MPUO) formalism and show how the effective symmetry operation on the boundary of CZX model can be expressed as MPUO. Moreover, we are going to relate MPUO of a symmetry group to the 3-cocycle of the group and in particular, we are going to show that the CZX model corresponds to a nontrivial 3-cocycle of the $Z_2$ group.

A matrix product operator acting on a 1D system is given by,\cite{MCP1012}
\be
O=\sum_{\{i_k\},\{i_k'\}}Tr(T^{i_1,i'_1}T^{i_2,i'_2}...T^{i_N,i'_N})|i'_1i'_2...i'_N\>\<i_1i_2...i_N|
\ee
where for fixed $i$ and $i'$, $T^{i,i'}$ is a matrix with index $\alpha$ and $\beta$. Here we want to use this formalism to study symmetry transformations, therefore we restrict $O$ to be a unitary operator $U$. Using matrix product representation, $U$ does not have to be an on-site symmetry. $U$ is represented by a rank-four tensor $T^{i,i'}_{\alpha,\beta}$ on each site, where $i$ and $i'$ are input and output physical indices and $\alpha$, $\beta$ are inner indices. Basic properties of matrix product unitary operators are given in appendix \ref{MPUO}.

In particular, the symmetry operator $U_{CZX}$ (we omit the $\sim$ label for effective spins in following discussions) on the boundary of the CZX model can be represented by tensors
\be
\begin{array}{l}
T^{0,1}(CZX) = |0\>\<+|, \\
T^{1,0}(CZX) = |1\>\<-|, \\
\text{other terms are zero}
\end{array}
\ee
where $|+\>=|0\>+|1\>$ and $|-\>=|0\>-|1\>$.  It is easy to check that this tensor indeed gives $U_{CZX}=CZ_{12}...CZ_{N1}X_1...X_N$.

The other element in the $Z_2$ group--the identity operation--can also be represented as MPUO with tensors
\be
\begin{array}{l}
T^{0,0}(I)=|0\>\<0|, \\
T^{1,1}(I)=|0\>\<0|, \\
\text{other terms are zero}
\end{array}
\ee
These two tensors are both in the canonical form as defined in appendix \ref{MPUO}.

\begin{figure}[ht]
\begin{center}
\includegraphics[scale=0.5]{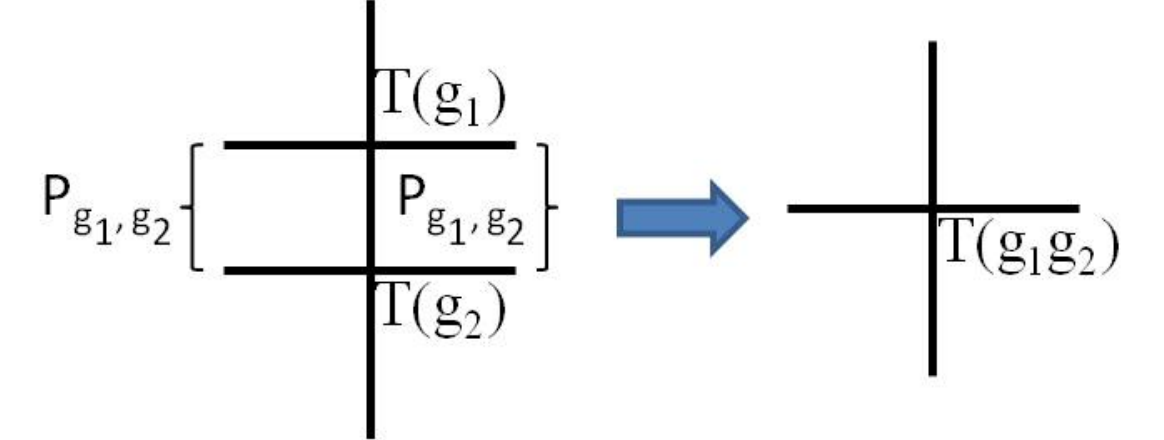}
\end{center}
\caption{Reduce combination of $T(g_2)$ and $T(g_1)$ into $T(g_1g_2)$. 
}
\label{P12}
\end{figure}

If two MPUO $T(g_2)$ and $T(g_1)$ are applied subsequently, their combined action should be equivalent to $T(g_1g_2)$. However, the tensor $T(g_1,g_2)$ obtained by contracting the output physical index of $T(g_2)$ with the input physical index of $T(g_1)$, see Fig. \ref{P12}, is usually more redundant than $T(g_1g_2)$ and might not be in the canonical form. It can only be reduced to $T(g_1g_2)$ if certain projection $P_{g_1,g_2}$ is applied to the inner indices (see Fig. \ref{P12}).

$P_{g_1,g_2}$ is only defined up to an arbitrary phase factor $e^{i\theta(g_1,g_2)}$. If the projection operator on the right side $P_{g_1,g_2}$ is changed by the phase factor $e^{i\theta(g_1,g_2)}$, the projection operator $P^{\dg}_{g_1,g_2}$ on the left side is changed by phase factor $e^{-i\theta(g_1,g_2)}$. Therefore the total action of $P_{g_1,g_2}$ and $P^{\dg}_{g_1,g_2}$ on $T(g_1,g_2)$ does not change and the reduction procedure illustrated in Fig.\ref{P12} still works. Moreover, from the discussion in the appendix \ref{MPUO}, we know that this is the only degree of freedom in $P_{g_1,g_2}$. Up to a phase factor, $P_{g_1,g_2}$ is unique (on the unique block in the canonical form of $T(g_1,g_2)$).

Let us illustrate how the reduction is done for the symmetry group $(I,U_{CZX})$. For example, if we apply $U_{CZX}U_{CZX}$ the totally action should be equivalent to $I$. However the tensor $T(CZX,CZX)$ is given by
\be
\begin{array}{l}
T^{0,0}(CZX,CZX)=|01\>\<+-|, \\
T^{1,1}(CZX,CZX)=|10\>\<-+|, \\
\text{other terms are zero}
\end{array}
\ee
This tensor is reduced to $T(I)$ if projection
\be
P_{CZX,CZX} = (|01\>-|10\>)\<0|
\ee
and its Hermitian conjugate are applied to the right and left of $T(CZX,CZX)$ respectively.\footnote{The mapping actually reduces $T(CZX,CZX)$ to $-T(I)$. But this is not a problem as we can redefine $\t{T}(CZX)=iT(CZX)$ and the extra minus sign would disappear.} Adding an arbitrary phase factor $e^{i\theta(CZX,CZX)}$ to $P_{CZX,CZX}$ does not affect the reduction at all. By writing $P_{CZX,CZX}$ in the above form, we have made a particular choice of phase. 

Below we list the (right) projection operators for all possible combinations of $g_1$ and $g_2$ of this $Z_2$ group.
\be
\begin{array}{lll}
P_{I,I} & = & |00\>\<0| \\
P_{CZX,I} & = & |00\>\<0|+|10\>\<1| \\
P_{I,CZX} & = & |00\>\<0|+|10\>\<1| \\
P_{CZX,CZX} & = & (|01\>-|10\>)\<0| \\
\end{array}
\ee
Note that in giving $P_{g_1,g_2}$ we have picked a particular choice of phase factor $e^{i\theta(g_1,g_2)}$. In general, any phase factor is allowed.

\begin{figure}[ht]
\begin{center}
\includegraphics[scale=0.5]{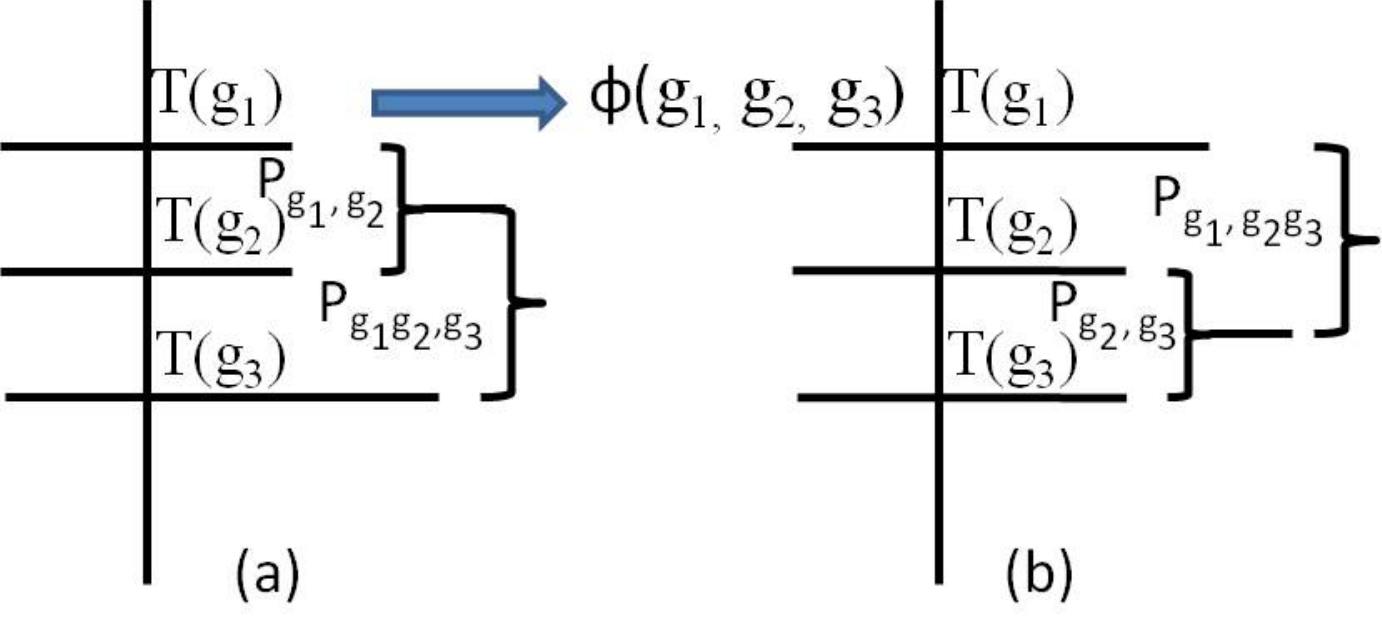}
\end{center}
\caption{Different ways to reduce combination of $T(g_3)$, $T(g_2)$ and $T(g_1)$ into $T(g_1g_2g_3)$. Only the right projection operators are shown. Their combined actions differ by a phase factor $\phi(g_1,g_2,g_3)$.
}
\label{P123}
\end{figure}

Nontrivial phase factors appear when we consider the combination of three MPUO's. See Fig. \ref{P123}. 

There are two different ways to reduce the tensors. We can either first reduce the combination of $T(g_1)$, $T(g_2)$ and then combine $T(g_3)$ or first reduce the combination of $T(g_2)$,$T(g_3)$ and then combine $T(g_1)$. The two different ways should be equivalent. More specifically, they should be the same up to phase on the unique block of $T{g_1,g_2,g_3}$. Denote the projection onto the unique block of $T(g_1,g_2,g_3)$ as $Q_{g_1,g_2,g_3}$. We find that
\be
\begin{array}{l}
Q_{g_1,g_2,g_3}(I_3\otimes P_{g_1,g_2})P_{g_1g_2,g_3}= \\
\phi(g_1,g_2,g_3) Q_{g_1,g_2,g_3}(P_{g_2,g_3}\otimes I_1)P_{g_1,g_2g_3}
\end{array}
\ee
From this we see that the reduction procedure is associative up to a phase factor $\phi(g_1,g_2,g_3)$. According to the definition of cocycles in appendix \ref{Gcoh}, we see that $\phi(g_1,g_2,g_3)$ forms a 3-cocycle of group $G$. That is, $\phi(g_1,g_2,g_3)$ satisfies
\be
\frac{ \phi(g_2,g_3,g_4) \phi(g_1,g_2g_3,g_4)\phi(g_1,g_2,g_3) }
{\phi(g_1g_2,g_3,g_4)\phi(g_1,g_2,g_3g_4)}
=1
\ee

Let's calculate $\phi(g_1,g_2,g_3)$ explicitly for the group generated by $U_{CZX}$.
\be
\begin{array}{ll}
\phi(I,I,I)=1 & \phi(I,I,CZX)=1 \\
\phi(I,CZX,I)=1 & \phi(CZX,I,I)=1 \\
\phi(I,CZX,CZX)=1 & \phi(CZX,CZX,I)=1\\
\phi(CZX,I,CZX)=1 & \phi(CZX,CZX,CZX)=-1
\end{array}
\ee
We can check that $\phi$ is indeed a 3-cocycle. The last term shows a nontrivial $-1$. This minus one cannot be removed by redefining the phase of $P_{g_1,g_2}$ in any way. Therefore $\phi$ corresponds to a nontrivial 3-cocycle for the $Z_2$ group.

What does this nontrivial mathematical structure imply about the physics of the CZX model? In the next section we are going to answer this question by proving that MPUO related to a nontrivial 3-cocycle cannot have a short range entangled symmetric state. That is, the boundary of the CZX model cannot have a gapped symmetric ground state. It either breaks the symmetry or is gapless.

\section{Nontrivial 3-cocycle of MPUO and nonexistence of SRE symmetric state}

In this section we will show that a symmetry defined by a MPUO on a 1D chain can have a SRE symmetric state only if the MPUO corresponds to a trivial 3-cocycle. Therefore, the boundary of the CZX model must be gapless or have symmetry breaking. For this proof, we will be using the matrix product state representation of SRE states.

Suppose that the symmetry on a 1D chain is represented by tensors $T^{i,i'}_{\alpha,\beta}(g)$. WLOG, $T(g)$ is single-blocked and in the canonical form as defined in appendix \ref{MPUO}. Assume that it has a SRE symmetric state represented by matrices $A^i_{\lambda,\eta}$ which is also single-blocked and in the canonical form. For a review of matrix product state formalism including its canonical form and single-block property see appendix \ref{MPS}.

\begin{figure}[ht]
\begin{center}
\includegraphics[scale=0.5]{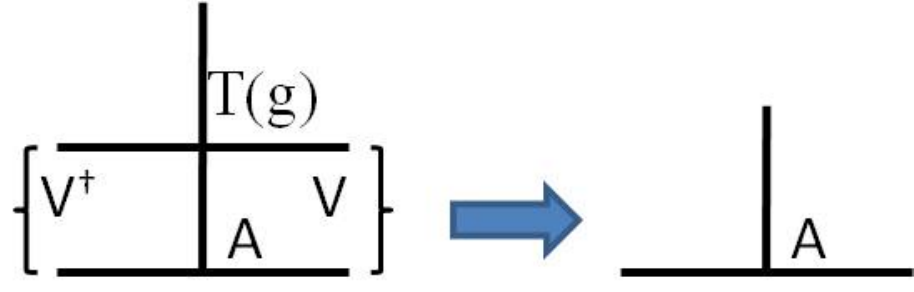}
\end{center}
\caption{Reduction of the combination of $T(g)$ and $A$ into $A$. Here $T^{i,i'}(g)$ is a MPUO, $A^i$ is a matrix product state symmetric under $T^{i,i'}(g)$.
}
\label{PgA}
\end{figure}

Based on the result in \Ref{PVW0701} and \Ref{PWS0802} we can show that (see appendix \ref{MPUO})
\be
A^i= V^{\dg} (\sum_{i'} T^{i,i'}(g)A^{i'}) V
\label{TA}
\ee
where $V^{\dg}V=I$ and $V$ is unique on the single block of $\sum_{i'} T^{i,i'}(g)A^{i'}$ up to phase. This is saying that we can reduce the MPS obtained from $\sum_{i'} T^{i,i'}(g)A^{i'}$ back to the original form $A^i$ by applying $V^{\dg}$ and $V$ to the left and right of the matrices respectively. See Fig. \ref{PgA}.

For a fixed representation of the SRE state $A^{i}$ and fixed representation of the MPUO symmetry $T(g)$, $V$ is fixed up to phase. We can pick a particular choice of phase for $V$.

\begin{figure}[ht]
\begin{center}
\includegraphics[scale=0.5]{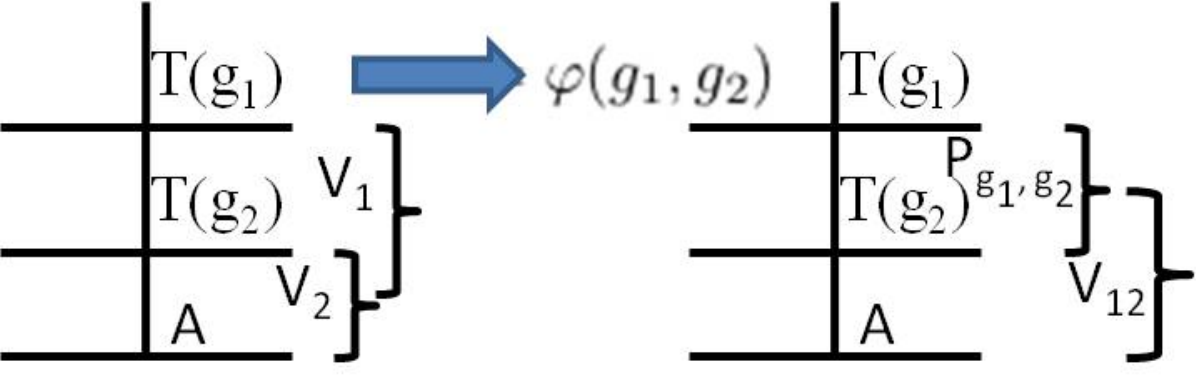}
\end{center}
\caption{Two ways of reducing the combination of $T(g_2)$, $T(g_1)$ and $A$ into $A$. Only the right projection operators are shown. Their combined actions differ by a phase factor $\varphi(g_1,g_2)$.
}
\label{Pg12A}
\end{figure}

Now we consider the combined operation of $T(g_1)$ and $T(g_2)$ on $A$. See Fig.\ref{Pg12A}.

We can either first combine $T(g_2)$ and $A$ and then combine $T(g_1)$ and $A$ or first combine $T(g_1)$ and $T(g_2)$ and then combine $T(g_1g_2)$ and $A$. The right projection operator for these two methods differ by a phase factor $\varphi(g_1,g_2)$. This phase factor can be arbitrarily changed by changing the phase of $P_{g_1,g_2}$. For following discussions, we fix the phase of $P_{g_1,g_2}$ and hence $\varphi(g_1,g_2)$.

\begin{figure}[ht]
\begin{center}
\includegraphics[scale=0.4]{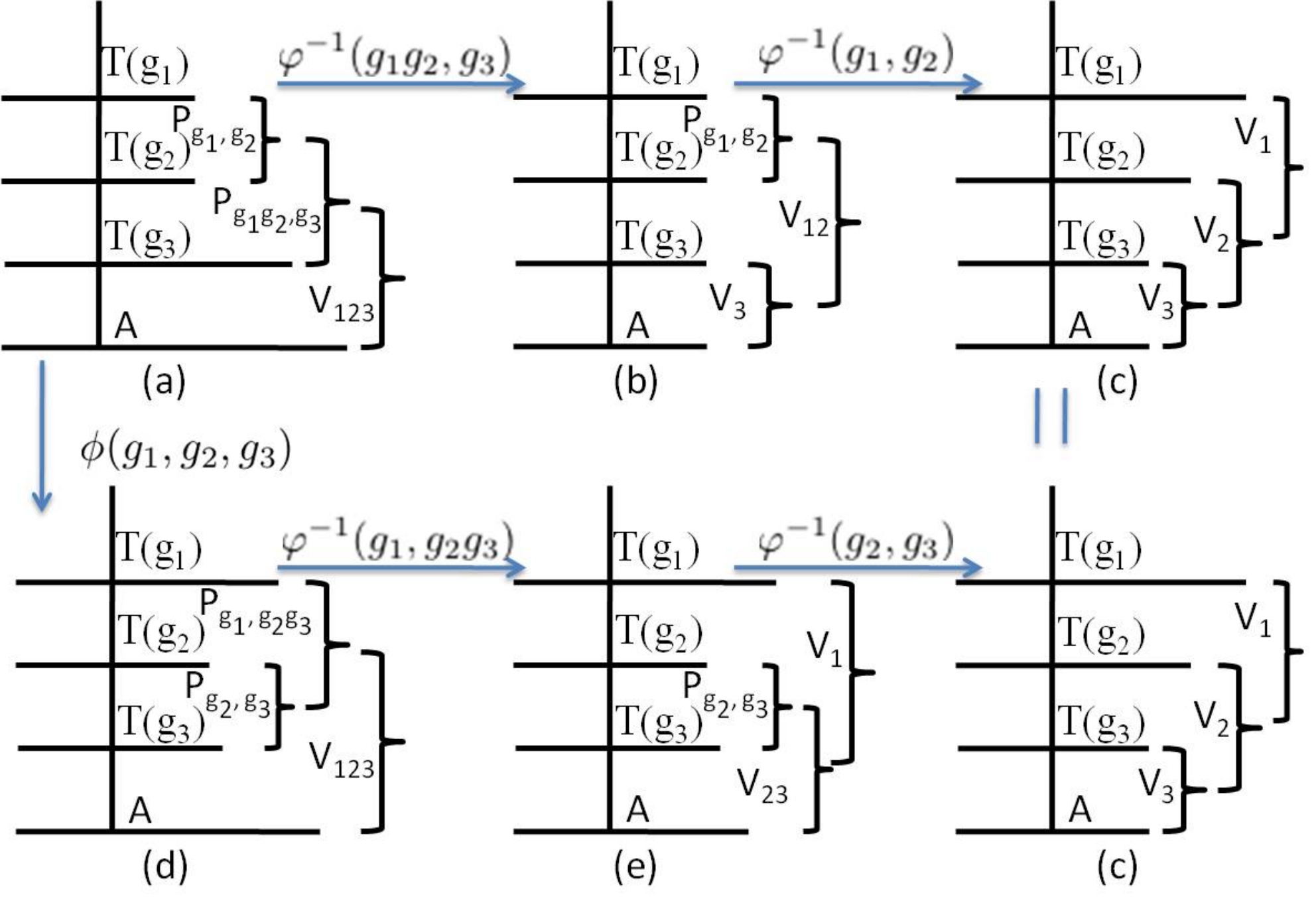}
\end{center}
\caption{Different ways of reducing the combination of $T(g_3)$, $T(g_2)$, $T(g_1)$ and $A$ into $A$. Only the right projection operators are shown. Their combined actions differ by a phase factor written on the arrow. 
}
\label{Pg123A}
\end{figure}

This is all the freedom we can have. If we are to combine three or more $T$'s with $A$, different reduction methods differ by a phase factor but the phase factor are all determined by $\varphi(g_1,g_2)$. Consider the situation in Fig. \ref{Pg123A}, where we are to combine $T(g_3)$, $T(g_2)$ and $T(g_1)$ with $A$. 

To change the reduction procedure in Fig.\ref{Pg123A}(a) to that in Fig.\ref{Pg123A}(c), we can either go through step (b) or steps (d) and (e). If we go through step (b), the phase difference in the right projection operators is
\be
\varphi^{-1}(g_1g_2,g_3)\varphi^{-1}(g_1,g_2)
\ee
On the other hand, if we go through steps (d) and (e), the phase difference in the right projection operators is
\be
\phi(g_1,g_2,g_3)\varphi^{-1}(g_1,g_2g_3)\varphi^{-1}(g_2,g_3)
\ee

But these two procedures should be equivalent as the initial and final configurations are the same whose phases have been fixed previously. Therefore, we find that
\be
\phi(g_1,g_2,g_3)=\frac{\varphi(g_1,g_2g_3)\varphi(g_2,g_3)}{\varphi(g_1g_2,g_3)\varphi(g_1,g_2)}
\ee
and $\phi(g_1,g_2,g_3)$ must be a trivial 3-cocycle (see Eq. \ref{3coboundary}). 

This finishes the proof that: {\it A 1D system with symmetry defined by matrix product unitary operators can have a gapped symmetric ground state only if the matrix product unitary operator corresponds to a trivial 3-cocycle.}

Because we have shown that the symmetry on the boundary of the CZX model corresponds to a nontrivial 3-cocycle of the $Z_2$ group, the system with boundary cannot have a gapped symmetric ground state. This shows that the CZX model has nontrivial SPT order protected by on-site $Z_2$ symmetry as we have promised in section \ref{CZX}.

\section{Generalization to fermion system}
\label{fermion}

Due to the interest in fermion SPT orders in interacting systems in two and higher dimensions, in this section we are going to give a fermionic version of the CZX model which also has nontrivial SPT order protected only by on-site $Z_2$ symmetry.

In constructing this model, first we identify each spin in the CZX model with a fermionic mode and the spin $|0\>$ state with the zero fermion state, the spin $|1\>$ state with the one fermion state. Each site then contains four modes (see Fig.\ref{CZX_model}).  Denote the creation and annihilation operator on each mode as $c^{\dg}_i$ and $c_i$. 

Fermion system has an intrinsic fermion parity symmetry which is an on-site $Z_2$ symmetry given by
\be
Pf=\prod^4_{i=1} (1-2c^{\dg}_ic_i)
\ee
This $Z_2$ symmetry is always preserved.

Similar to the CZX model we define another on-site $Z_2$ symmetry $U^f_{CZX}$, which is going to protect the nontrivial SPT order. 
\be
U^f_{CZX}=U^f_XU^f_{CZ}
\ee
where
\be
U^f_X=\prod^4_{i=1} (c^{\dg}_i+c_i)
\ee
is a particle-hole transformation and
\be
U^f_{CZ}=\prod^4_{i=1} (I-2c^{\dg}_ic_ic^{\dg}_{i+1}c_{i+1})
\ee
It can be checked that $U^f_X$ and $U^f_{CZ}$ commute with each other and they both commute with $Pf$. Therefore $U^f_{CZX}$ commutes with $Pf$. $U^f_{CZX}$ generates an on-site $Z_2$ symmetry. 

The Hamiltonian of the system is again a sum of local terms around each plaquette. $H^f=\sum H^f_{p_i}$.
\be
H^f_{p_i}=-X^f_4\otimes P_2^{u,f}\otimes P_2^{d,f}\otimes P_2^{l,f}\otimes P_2^{r,f}
\ee
(see Fig.\ref{CZX_model}(c))where $X^f_4$ acts on the four modes in the middle plaquette as
\be
X^f_4=c_4c_3c_2c_1+c^{\dg}_1c^{\dg}_2c^{\dg}_3c^{\dg}_4
\ee
and $P^f_2$ acts on the two modes in every half plaquette as
\be
P^f_2=c_ic^{\dg}_ic_{i+1}c^{\dg}_{i+1}+c^{\dg}_ic_ic^{\dg}_{i+1}c_{i+1}
\ee
For the remaining four modes at the corner, $H^f_{p_i}$ acts as identity on them. It can be checked that the Hamiltonian satisfies the fermion parity symmetry and the on-site $Z_2$ symmetry generated by $U^f_{CZX}$. Moreover terms around different plaquettes commute with each other.

The ground state is then a product of plaquette states
\be
|\psi^f_{p_i}\>=(1+c^{\dg}_1c^{\dg}_2c^{\dg}_3c^{\dg}_4)|\Omega\>
\ee
where $|\Omega\>$ is vacuum state on the four modes $1 \sim 4$ around a plaquette. The ground state is short range entangled and symmetric under both $Pf$ and $U^f_{CZX}$. If $U^f_{CZX}$ can be violated, we can disentangle this state into a product of states on each mode without violating the fermion parity symmetry. However, if $U^f_{CZX}$ is preserved, this state is inequivalent from a trivial product state.

The nontrivial-ness of this model can be seen again from the boundary. The boundary of this fermion model is the same as that of the spin model, because the effective degrees of freedom in each half plaquette has two states: the vacuum state on the two modes and the fully occupied state on the two modes. These two states are both bosonic, therefore the boundary can be treated as a spin system just like for the CZX model. The symmetry on the boundary is again generated by $U_{CZX}$ which we have shown cannot have a SRE symmetric state. Therefore, this fermionic CZX model has a nontrivial on-site $Z_2$ symmetry protected SPT order.

\section{Summary}

In this paper, we have given the explicit construction of a two dimensional interacting spin model with nontrivial on-site $Z_2$ symmetry protected topological order. We found that the system is highly nontrivial because if it has a boundary the boundary is either gapless or breaks symmetry. We showed this by writing the effective symmetry transformation on the boundary as a matrix product unitary operator and revealed a nontrivial 3-cocycle structure in its transformation rule. We proved that any matrix product unitary operator related to a nontrivial 3-cocycle in $\cH^3(G,U(1))$ cannot have a gapped short range entangled symmetric state.

This model could have interesting implication for the study of topological phases using tensor network presentation. In the tensor network repretation of topological phases, it has been understood that in one dimension injective tensors provide a complete characterization of gapped ground states and its gauge transformation under symmetry reals the SPT order of the phase.\cite{SPC1032,CGW1107} In higher dimensions, similar analysis of injective tensors have been carried out.\cite{PVC0850,PSG1010,SCP1053} However, the fact that the ground state wavefunction of CZX model has a loop structure and cannot be represented by an injective tensor tells us that we need to consider more general forms of tensors in order to study interesting SPT orders in more than one dimension. Identifying the proper set of tensors for the characterization of gapped short range entangled phases in higher dimensions is an important open question. Or an alternative approach is to reduce the problem from 2D to 1D by considering the tensor representation of effective symmetry transformation on the boundary, as was done in this paper. How the reduction can be done in more than two dimension is unknown.

The 1D boundary of the CZX model presents new challenges to our understanding of 1D systems. While it is a locally interacting system with $Z_2$ symmetry, it does not have a gapped symmetric phase like in transverse Ising model. Moreover, the gapless excitations cannot be gapped by breaking translational symmetry like in spin $1/2$ chains. The peculiarity of this system originates from the fact that this 1D system can only exist as the boundary of a 2D system and not on its own. Finding a proper field theory description of this system would expand our current understanding of 1D physics.

The relation between SPT order and cocycle is not accidental. Actually the pattern has shown up in lower dimensions.\cite{SPC1032} In zero dimension, symmetric states are classified by 1D representation of the group, that is, class of 1-cocycles in $\cH^1(G,U(1))$ and in one dimension SPT phases are classified by projective representations of the group, that is, class of 2-cocycles in $\cH^2(G,U(1))$. Here we make a connection between 2D SPT order and class of 3-cocycles in $\cH^3(G,U(1))$. In fact, this relation is more general. In another paper, we are going to show that actually $d$ dimensional SPT orders are related to $(d+1)$-cocycles in $\cH^{d+1}(G,U(1))$, which could lead to a full classification of SPT orders in any dimension.

We would like to thank Zhenghan Wang for helpful discussions.
This research is supported by NSF Grant No. DMR-1005541 and NSFC 11074140. 

\appendix

\section{Projective Representation} 
\label{prorep}

Matrices $u(g)$ form a projective representation of symmetry group $G$ if
\begin{align}
 u(g_1)u(g_2)=\om(g_1,g_2)u(g_1g_2),\ \ \ \ \
g_1,g_2\in G.
\end{align}
Here $\om(g_1,g_2) \in U(1)$ and $\om(g_1,g_2) \neq 0$, which is called the
factor system of the projective representation. The factor system satisfies
\begin{align}
 \om(g_2,g_3)\om(g_1,g_2g_3)&=
 \om(g_1,g_2)\om(g_1g_2,g_3),
\end{align}
for all $g_1,g_2,g_3\in G$.
If $\om(g_1,g_2)=1$, this reduces to the usual linear representation of $G$.

A different choice of pre-factor for the representation matrices
$u'(g)= \bt(g) u(g)$ will lead to a different factor system
$\om'(g_1,g_2)$:
\begin{align}
\label{omom}
 \om'(g_1,g_2) =
\frac{\bt(g_1g_2)}{\bt(g_1)\bt(g_2)}
 \om(g_1,g_2).
\end{align}
We regard $u'(g)$ and $u(g)$ that differ only by a pre-factor as equivalent
projective representations and the corresponding factor systems $\om'(g_1,g_2)$
and $\om(g_1,g_2)$ as belonging to the same class $\om$.

Suppose that we have one projective representation $u_1(g)$ with factor system
$\om_1(g_1,g_2)$ of class $\om_1$ and another $u_2(g)$ with factor system
$\om_2(g_1,g_2)$ of class $\om_2$, obviously $u_1(g)\otimes u_2(g)$ is a
projective presentation with factor group $\om_1(g_1,g_2)\om_2(g_1,g_2)$. The
corresponding class $\om$ can be written as a sum $\om_1+\om_2$. Under such an
addition rule, the equivalence classes of factor systems form an Abelian group,
which is called the second cohomology group of $G$ and denoted as
$\cH^2[G,U(1)]$.  The identity element $1 \in \cH^2[G,U(1)]$ is the class that
corresponds to the linear representation of the group.

\section{Group cohomology }
\label{Gcoh}

The above discussion on the factor system of a projective representation can be
generalized which give rise to a cohomology theory of group.  In this section,
we will briefly describe the group cohomology theory.

For a group $G$, let $M$ be a G-module, which is an abelian group (with
multiplication operation) on which $G$ acts compatibly with the multiplication
operation (\ie the abelian group structure) on M:
\begin{align}
\label{gm}
 g\cdot (ab)=(g\cdot a)(g\cdot b),\ \ \ \ g\in G,\ \ \ \ a,b\in M.
\end{align}
For the cases studied in this paper, $M$ is simply the $U(1)$ group and $a$ an
$U(1)$ phase.  The multiplication operation $ab$ is the usual multiplication of
the $U(1)$ phases.  The group action is trivial: $g\cdot a=a$, $g\in G$, $a=\in
U(1)$.

Let $\om_n(g_1,...,g_n)$ be a function of $n$ group
elements whose value is in the G-module $M$. In other words, $\om_n:
G^n\to M$.  Let $\cC^n(G,M)=\{\om_n \}$ be the space of all such
functions.  
Note that $\cC^n(G,M)$ is an Abelian group
under the function multiplication 
$ \om''_n(g_1,...,g_n)= \om_n(g_1,...,g_n) \om'_n(g_1,...,g_n) $.
We define a map $d_n$ from $\cC^n[G,U(1)]$ to $\cC^{n+1}[G,U(1)]$:
\begin{align}
&\ \ \ \
(d_n \om_n) (g_1,...,g_{n+1})=
\nonumber\\
&
g_1\cdot \om_n (g_2,...,g_{n+1})
\om_n^{(-1)^{n+1}} (g_1,...,g_{n}) \times
\nonumber\\
&\ \ \ \ \
\prod_{i=1}^n
\om_n^{(-1)^i} (g_1,...,g_{i-1},g_ig_{i+1},g_{i+2},...g_{n+1})
\end{align}
Let
\begin{align}
 \cB^n(G,M)=\{ \om_n| \om_n=d_{n-1} \om_{n-1}|  \om_{n-1} \in \cC^{n-1}(G,M) \}
\end{align}
and
\begin{align}
 \cZ^n(G,M)=\{ \om_{n}|d_n \om_n=1,  \om_{n} \in \cC^{n}(G,M) \}
\end{align}
$\cB^n(G,M)$ and $\cZ^n(G,M)$ are also Abelian groups
which satisfy $\cB^n(G,M) \subset \cZ^n(G,M)$ where
$\cB^1(G,M)\equiv \{ 1\}$.
The $n$-cocycle of $G$ is defined as
\begin{align}
 \cH^n(G,M)= \cZ^n(G,M) /\cB^n(G,M) 
\end{align}

Let us discuss some examples. 
We choose $M=U(1)$ and $G$ acts trivially: $g\cdot a=a$, $g\in G$, $a\in U(1)$.
In this case $\om_n(g_1,...,g_n)$ is just a phase factor.
From
\begin{align}
 (d_1 \om_1)(g_1,g_2)= \om_1(g_2)\om_1(g_1)/\om_1(g_1g_2)
\end{align}
we see that
\begin{align}
 \cZ^1(G,U(1))=\{  \om_1| \om_1(g_2)\om_1(g_1)=\om_1(g_1g_2) \} .
\end{align}
In other words, $\cZ^1(G,U(1))$ is the set formed by all the 1D representations
of $G$.  Since $\cB^1(G,U(1))\equiv \{ 1\}$ is trival.
$\cH^1(G,U(1))=\cZ^1(G,U(1))$ is also the set of all the 1D representations of
$G$.

From
\begin{align}
&\ \ \ \ (d_2 \om_2)(g_1,g_2,g_3)
\nonumber\\
&= 
\om_2(g_2,g_3) \om_2(g_1,g_2g_3)/\om_2(g_1g_2,g_3)\om_2(g_1,g_2)
\end{align}
we see that
\begin{align}
& \cZ^2(G,U(1))=\{  \om_2| 
\\
&\ \ \ \om_2(g_2,g_3) \om_2(g_1,g_2g_3)=\om_2(g_1g_2,g_3)\om_2(g_1,g_2)
 \} .
\nonumber 
\end{align}
and
\begin{align}
& \cB^2(G,U(1))=\{ \om_2|\om_2(g_1,g_2)=\om_1(g_2)\om_1(g_1)/\om_1(g_1g_2)
 \} .
\end{align}
The 2-cocycle
$\cH^2(G,U(1))=\cZ^2(G,U(1))/\cB^2(G,U(1))$ classify the
projective representations discussed in section \ref{prorep}.

From
\begin{align}
&\ \ \ \ (d_3 \om_3)(g_1,g_2,g_3,g_4)
\nonumber\\
&= \frac{ \om_3(g_2,g_3,g_4) \om_3(g_1,g_2g_3,g_4)\om_3(g_1,g_2,g_3) }
{\om_3(g_1g_2,g_3,g_4)\om_3(g_1,g_2,g_3g_4)}
\end{align}
we see that
\begin{align}
& \cZ^3(G,U(1))=\{  \om_3| 
\\
&\ \ \ \frac{ \om_3(g_2,g_3,g_4) \om_3(g_1,g_2g_3,g_4)\om_3(g_1,g_2,g_3) }
{\om_3(g_1g_2,g_3,g_4)\om_3(g_1,g_2,g_3g_4)}
=1
 \} .
\nonumber 
\end{align}
and
\begin{align}
& \cB^3(G,U(1))=\{ \om_3| \om_3(g_1,g_2,g_3)=\frac{
\om_2(g_2,g_3) \om_2(g_1,g_2g_3)}{\om_2(g_1g_2,g_3)\om_2(g_1,g_2)}
 \},
\label{3coboundary}
\end{align}
which give us the 3-cocycle
$\cH^3(G,U(1))=\cZ^3(G,U(1))/\cB^3(G,U(1))$.

\section{Review: matrix product states and its canonical form}
\label{MPS}

In this section we review matrix product state and its canonical form which was first derived in \Ref{PVW0701}. Similar ideas are going to be used in the study of matrix product unitary operators.

A matrix product representation of 1D state is
\be
|\psi\>=\sum_{i_1i_2...i_N}Tr(A_{i_1}A_{i_2}...A_{i_N})|i_1i_2...i_N\>
\ee
$A_i$'s are $D\times D$ matrices.

Define double tensor $E$ for the MPS as
\be
E=\sum_i A_i \otimes A_i^*
\ee
Equivalently, $E$ can be expressed as a completely positive quantum channel $\mathcal{E}$ as
\be
\mathcal{E}(X)=\sum_i A_i X A^{\dg}_i
\ee
and the corresponding dual channel $\mathcal{E}^*$ as
\be
\mathcal{E}^*(X)=\sum_i A^{\dg}_i X A_i
\ee

The correspondence between $E$ and $\mathcal{E}$, $\mathcal{E}^*$ is as follows. Suppose that $X$ and $Y$ are $D\times D$ matrices which satisfy
\be
Y=\mathcal{E}(X)
\ee
Combine the two indices of the matrices into one and write them as vectors
\be
(V_X)_{(\alpha-1)D+\beta} = X_{\alpha,\beta}\, \  (V_Y)_{(\alpha-1)D+\beta} = Y_{\alpha,\beta}
\ee
$V_X$ and $V_Y$ are then related by $E$ as
\be
EV_X=V_Y
\ee
Similarly, if
\be
Y=\mathcal{E}^*(X)
\ee
then
\be
V_X^{\dg}E=V_Y^{\dg}
\ee
We will use $E$ and $\mathcal{E}$, $\mathcal{E}^*$ inter-changably, whichever is more convenient.

From the structure of $\mathcal{E}$ and $\mathcal{E}^*$ we can put $A_i$'s into a canonical form. Suppose that the largest magnitude of the eigenvalues of $\mathcal{E}$ is $\lambda_1>0$. There could be multiple eigenvalues  $\lambda_1 e^{i\theta_k}$ of this magnitude.  As shown in \Ref{FNW9243}, $\e^{i\theta_k}$ form a group and they are the $p$th root of unity. To get rid of this, we can just group $p$ sites together and the eigenvalues of the largest magnitude will all be real and positive. We still label them as $\lambda_1$. 

Because $\mathcal{E}$ is a completely positive channel, at least one of corresponding fixed points $\Lambda$ 
\be
\mathcal{E}(\Lambda)=\lambda_1\Lambda
\ee
is positive-semidefinite. Denote the support space of $\Lambda$ as $P$. It can be shown that $A_iP=PA_iP$.\cite{PVW0701} Decompose each $A_i$ into four parts $A_i=PA_iP+PA_iP_{\perp}+P_{\perp}A_iP+P_{\perp}A_iP_{\perp}$.  $P_{\perp}A_iP=0$. $PA_iP_{\perp}$ may not be zero. However, it does not contribute to the MPS, therefore we can remove it safely. After doing this, $A_i$ is decomposed into two blocks and $\Lambda$ is a full rank positive fixed point of $\mathcal{E}_P(X)=\sum_i (PA_iP) X (PA_iP)^{\dg}$ with eigenvalue $\lambda_1$.

Because 
\be
\mathcal{E}_P(X)=\sum_i (PA_iP) X (PA_iP)^{\dg}=\sum_i (A_iP) X (A_iP)^{\dg}
\ee
every fixed point of $\mathcal{E}_P$(within space $P$) is also a fixed point of $\mathcal{E}$ with the same eigenvalue. Therefore, $\lambda_1$ is also the largest eigenvalue of $\mathcal{E}_P$. Suppose that $\mathcal{E}_P$ has another fixed point $Z$ of eigenvalue $\lambda_1$ which is not proportional to $\Lambda$. WLOG, we can choose $Z$ to be Hermitian. (This is because $\sum_i (A_iP) Z (A_iP)^{\dg}=\lambda_1 Z$, therefore $\sum_i (A_iP) Z^{\dg} (A_iP)^{\dg}=\lambda_1 Z^{\dg}$. And because $Z$ is not proportional to $\Lambda$, at least one of the Hermitian matrices $Z+Z^{\dg}$ or $i(Z-Z^{\dg})$ is not proportional to $\Lambda$.) Diagonalize the Hermitian matrix $\Lambda^{-1/2}Z\Lambda^{-1/2}$ and get eigenvalues $z_1>z_2>$... It is easy to see that $\Lambda-\frac{1}{z_1}Z$ is another non full rank positive fixed point of $\mathcal{E}_P$ with eigenvalue $\lambda_1$. Therefore we can repeat the previous process and turn $PA_iP$ into smaller blocks.

Repeat this process for every block until (1) the channel $\mathcal{E}_{P_k}$ of every block $k$ has a largest positive eigenvalue $\lambda_k$. There is a positive full rank fixed point $\Lambda_{P_k}$ within subspace $P_k$. (2) There is no other fixed point within $P_k$ of the same eigenvalue. (3) The block $P_{\perp}=I-\sum_k P_k$ which does not have a positive fixed point for largest eigenvalue must only have zero eigenvalue. The block could be non-zero in general, but it does not contribute to MPS. Note that $\sum_k P_k + P_{\perp}=I$, $A_iP_k=P_kA_iP_k$. Written in the blocks $P_k$ and $P_{\perp}$, $A_i$ is upper(or lower) triangular. 

Now we look at each block $k$ separately but from the dual channel perspective. We can similarly block diagonalize $A_i^k$ if non full rank positive fixed point exists for the largest eigenvalue of $\mathcal{E}^*_{P_k}$. For each sub-block projection $P_{k,l}$, $P_{k,l}A_i^k=P_{k,l}A_i^kP_{k,l}$. $A_i^k$ can be turned into sub-blocks $A_i^{k,l}=P_{k,l}A_i^kP_{k,l}$. Note that, if $\Lambda_{P_{k,l}}=P_{k,l}\Lambda_{P_{k}}P_{k,l}$,
\be
\begin{array}{lll}
\sum_i A_i^{k,l}\Lambda_{P_{k,l}}(A_i^{k,l})^{\dg} & = & A_i^{k,l}\Lambda_{P_{k}}(A_i^{k,l})^{\dg}\\
 & = & P_{k,l}A_i^k \Lambda_{P_{k}}(A_i^k)^{\dg}P_{k,l} \\
 & = & \lambda_k \Lambda_{P_{k,l}}
\end{array}
\ee
Therefore, within each sub-block, $\Lambda_{P_{k,l}}$ is still a positive full rank fixed point of $\mathcal{E}_{P_{k,l}}$ with eigenvalue $\lambda_k$. As there cannot be positive fixed points of other eigenvalue, $\lambda_k$ must be the largest. Similarly, if $X_k$ is a fixed point of $\mathcal{E}_{P_k}$, $P_{k,l}X_kP_{k,l}$ is a fixed point of $\mathcal{E}_{P_{k,l}}$ with the same eigenvalue. 

Proceed similarly as for $\mathcal{E}$, we can block diagonalize $A_i^k$ into $A_i^{k,l}$ such that $\mathcal{E}^*_{P_{k,l}}$ has only one fixed point for its largest eigenvalue which is full rank positive.

Finally, we arrive at a canonical form, which is composed of blocks $P_k$ and sub-blocks $P_{k,l}$. Within each sub-block, the matrices satisfy (1) the channel $\mathcal{E}_{P_{k,l}}$ has a largest positive eigenvalue. The corresponding fixed point is full rank positive. (2) There is no other fixed point within the sub-block of the same eigenvalue. (3) the dual channel $\mathcal{E}^*_{P_{k,l}}$ also has a largest positive eigenvalue. The corresponding fixed point is full rank positive. (4) There is no other fixed point within the sub-block of the same eigenvalue. 

A generic matrix product state has only one block in its canonical form.\cite{PVW0701} We will call these MPS single-blocked MPS. Single-blocked MPS represents gapped, short range correlated 1D states. The single-block property is a generalization of the injectivity condition for MPS.\cite{PVW0701} A single-blocked MPS is injective if the dimension of the matrices equals that in the canonical form. On the other hand, a single-blocked MPS might not be written in a canonical form. It is in general more redundant. To do the reduction, necessary steps involves projection onto the single block and re-labeling the basis. Any invertible operation within the projected space might be added. However, if the resulting canonical form is fixed, the reduction operation is unique within the projected space up to an arbitrary phase factor.  

\section{Matrix Product Unitary Operators}
\label{MPUO}

Similarly to MPS, a matrix product representation of operators acting on a 1D system is given by,\cite{MCP1012}
\be
O=\sum_{\{i_k\},\{i_k'\}}Tr(T^{i_1,i'_1}T^{i_2,i'_2}...T^{i_N,i'_N})|i'_1i'_2...i'_N\>\<i_1i_2...i_N|
\ee

Here we restrict to unitary operators $U$ as we want to discuss symmetry operations. Using matrix product representation, $U$ does not have to be an on-site symmetry. $U$ is represented by a rank-four tensor $T^{i,i'}_{\alpha,\beta}$ on each site, where $i$ and $i'$ are input and output physical indices and $\alpha$, $\beta$ are inner indices.

Just like every matrix product state can be reduced to a canonical form.\cite{PVW0701} every matrix product operator can be reduced to a canonical form also. To do so, we just need to treat the two physical indices as one and apply the procedure described in appendix \ref{MPS}. Similar to MPS, we can also define double tensor/ quantum channel for each matrix product operator. The double tensor of $T$ is
\be
E=\sum_{i,i'} T^{i,i'}\otimes (T^{i,i'})^*
\ee 

The fact that $T$ represents a unitary operator puts strong constraint on the form of $T$. $U^{\dg}U=I\otimes...\otimes I$ is represented on each site by tensor
\be
\mathbb{T}^{i,i''}_{\alpha\alpha',\beta\beta'}=\sum_{i'} T^{i,i'}_{\alpha,\beta}(T^{i'',i'}_{\alpha',\beta'})^*
\ee
$\mathbb{T}$ must be equivalent to $\delta_{i,i''}$ on each site.  We can reduce $\mathbb{T}$ to the canonical form. The canonical form of $\mathbb{T}$ could contain multiple blocks, but each block must represent the same operator $I\otimes...\otimes I$ and takes the form $\lambda_k\delta_{i,i''}|k\>\<k|$. $|k\>\<k|$ is the projection onto the $k$th block, $\lambda_k$ is a number. Later we will impose further constraints on $U$ to get rid of multi-block.

First we want to show that we can write every MPUO in an single-blocked canonical form. That is, the canonical form contains only one block.
Suppose that we start with a canonical representation of the symmetry operation. In general, the canonical representation could have multiple blocks. We are going to show that this is not necessary as different blocks represent the same unitary operation.

Suppose that a canonical MPUO contains two blocks
\be
T^{ii'}=T_{[1]}^{ii'}\oplus T_{[2]}^{ii'}
\ee
$T_{[1]}$ represents MPO $O_1$ and $T_{[2]}$ represents MPO $O_2$ (not necessarily unitary). $U=O_1+O_2$. 

The corresponding $\mathbb{T}$ contains four blocks
\be
\begin{array}{lll}
\mathbb{T}^{i,i''} & = & \sum_{i'} T^{ii'}\otimes(T^{i''i'})^* \\
 & = & \mathbb{T}^{i,i'}_{[11]} \oplus \mathbb{T}^{i,i'}_{[12]} \oplus \mathbb{T}^{i,i'}_{[21]} \oplus \mathbb{T}^{i,i'}_{[22]}
\end{array}
\ee
$\mathbb{T}_{[kk']}$ represent MPO $O_kO_{k'}^{\dg}$. Because $\mathbb{T}$ represents $I\otimes I...\otimes I$, each of its block must also do. Therefore, 
\be
O_1O_1^{\dg}=O_1O_2^{\dg}=O_2O_1^{\dg}=O_2O_2^{\dg}=I \otimes I...\otimes I
\ee
That is, $O_1$ and $O_2$ represent the same unitary operator and there is no need for multiple blocks. In the following we will always assume that $T$ is written in a canonical form with only one block. We will call this the single-block condition for MPUO.

With the MPUO representation defined for each symmetry operation, we now want to know how the representation changes when two or more operations are combined. 

First let's consider what happens when $U$ is combined with $U^{\dg}$. As we discussed before, this is represented by $\mathbb{T}$ which could contain multiple blocks $\lambda_k\delta_{i,i''}|k\>\<k|$ in the canonical form. Correspondingly, the double tensor of $T$
\be
E=\sum_{i,i'} T^{i,i'}\otimes (T^{i,i'})^*=\sum_i \mathbb{T}^{i,i}
\ee
has multiple eigenvectors $|k\>$ with corresponding eigenvalues $\lambda_k$.

Define the correlator between two sets of operator pairs $\{o_1^m,\t{o}_1^m\}$ and $\{o_2^n,\t{o}_2^n\}$ to be
\be
\begin{array}{lll}
(o_1,o_2)_U & = & \sum_{mn} \Tr(o_1^mo_2^nU\t{o}_1^m\t{o}_2^nU^{\dg}) \\
            &   & - (\sum_m \Tr(o_1^mU\t{o}_1^mU^{\dg}))(\sum_n \Tr(o_2^nU\t{o}_2^nU^{\dg}))
\end{array}
\ee
On the one hand, written in terms of tensors, the correlator is expressed as
\be
\begin{array}{lll}
(o_1,o_2)_U & = & \Tr(E..E[o_1]..E[o_2]..E) \\
                     & - &\Tr(E..E[o_1]..E)\Tr(E..E[o_2]..E)
\end{array}
\ee 
where 
\be
\begin{array}{l}
E_{[o_1]}=\sum_{m,i} (o_1^m)^{i_2,i_3}(\t{o}_1^m)^{i_4,i_1}T^{i_1,i_2}\otimes(T^{i_4,i_3})^*\\
E_{[o_2]}=\sum_{n,i} (o_2^n)^{i_2,i_3}(\t{o}_2^n)^{i_4,i_1}T^{i_1,i_2}\otimes(T^{i_4,i_3})^*
\end{array}
\ee
This is the same form as the correlation function of operators $o_1=\sum_m o_1^m\otimes \t{o}_1^m$ and $o_2=\sum_n o_2^n\otimes \t{o}_2^n$ in a matrix product state with double tensor $E$. From our knowledge of MPS, we know that the correlator decays as $(\lambda_2/\lambda_1)^l$.

On the other hand, we consider for simplicity only unitaries $U$ which preserve locality of operators exactly. That is, if $o$ is supported on a finite number of sites, $UoU^{\dg}$ is also supported on a finite number of sites, though the number may be larger. We do not consider the local operators with exponentially decaying tails. \footnote{This is a reasonable restriction because we want to study systems at fixed point where the correlation length in the bulk is zero. Any exactly local operator in the bulk becomes an exactly local effective operator on the boundary.} Under this restriction, it follows that when $\{o_1^m,\t{o}_1^m\}$ and $\{o_2^n,\t{o}_2^n\}$ are far apart
\be
\begin{array}{lll}
\sum_{mn} \Tr(o_1^mo_2^nU\t{o}_1^m\t{o}_2^nU^{\dg}) \\
 = \sum_{mn} \Tr(o_1^mo_2^nU\t{o}_1^mUU^{\dg}\t{o}_2^nU^{\dg}) \\
 = \sum_{mn} \Tr((o_1^mU\t{o}_1^mU^{\dg}))\otimes(o_2^nU\t{o}_2^nU^{\dg})) \\
 = \sum_{mn} \Tr(o_1^mU\t{o}_1^mU^{\dg})\Tr(o_2^nU\t{o}_2^nU^{\dg})
\end{array}
\ee
the correlator $(o_1,o_2)_U$ must be zero if the separation is large enough. Therefore, $\lambda_2=0$. $E$ has only one eigenvector and $\mathbb{T}$ has only one block in its canonical form.

Now we want to use this property to show that the single-block condition is stable under combination of MPUO's. That is, if we start with two MPUO represented by $T^a$ and $T^b$ with only one block in the canonical form, their combination
\be
T^{c,ii''}_{\alpha\alpha',\beta\beta'}=\sum_{i'} T^{a,ii'}_{\alpha,\beta}T^{b,i'i''}_{\alpha',\beta'}
\ee
also has only one block in its canonical form. Of course, written as above, $T^c$ is not necessarily in the canonical form. Note that the discussion in the previous paragraphs is actually on the special case where $T^{b,ii'}=(T^{a,i'i})^*$.

In order to see this, we take the double tensor of $T^c$
\be
\begin{array}{lll}
E^{c} & = & \sum_{i,i''} T^{c,ii''}\otimes (T^{c,ii''})^* \\
    & = & \sum_{i,i''} (\sum_{i'_1}T^{a,ii'_1}\otimes T^{b,i'_1i''}) \otimes (\sum_{i'_2}T^{a,ii'_2}\otimes T^{b,i'_2i''})^* \\
    & = & \sum_{i'_1,i'_2} \mathbb{T}^{a,i'_1i'_2} \otimes \mathbb{T}^{b,i'_1i'_2}
\end{array}
\ee
$\mathbb{T}^a$ and $\mathbb{T}^b$ both have one block in their canonical form. Denote the projection onto the blocks as $P_a$ and $P_b$.
\be
\begin{array}{l}
P_a=|\psi^a_{\alpha\t{\alpha}}\>\<\psi^a_{\beta\t{\beta}}| \\
P_b=|\psi^b_{\alpha'\t{\alpha}'}\>\<\psi^b_{\beta'\t{\beta}'}|
\end{array}
\ee
Being the only eigenvector of $E^a$ and $E^b$, $|\psi^a_{\alpha\t{\alpha}}\>$ and $|\psi^b_{\alpha'\t{\alpha}'}\>$ are positive full rank if written as matrices $\Lambda^a_{\alpha,\t{\alpha}}$, $\Lambda^b_{\alpha',\t{\alpha}'}$. The only term that contributes to the trace of $E^c$ is
\be
(|\psi^a\> \otimes|\psi^b\>)(\<\psi^a| \otimes\<\psi^b|)
\ee
This is also true for any power of $E^c$.

This special property of $E^c$ tells us that $E^c$ has only a single non-zero eigenvalue. Suppose $E^c=\lambda_0|0\>\<0|+M$, $|0\>$ is short for $|\psi^a\> \otimes|\psi^b\>$. $\Tr(E^c)=\lambda_0$. Moreover, $\Tr(E^c)^k=\lambda_0^k$. Because $\Tr(E^c)^k=\sum_i (\lambda_i)^k$, it can be shown that $\lambda_i=0$, $\forall i>0$. The fact that $E^c$ has a single eigenvalue in turn tells us that $T^c$ contains only one block in its canonical form, because otherwise, $E^c$ would have at least $n^2$ non-zero eigenvalues with $n$ being the block number.

Therefore, we have shown that if we start with the canonical single-blocked tensor representation of some unitary operators, the tensor obtained from their concatenation still has only one block in its canonical form and is hence single-blocked. For single-blocked $T$ we can always apply the procedure in \Ref{PVW0701} (also discussed in detail in appendix \ref{MPS}) to reduce it to a canonical form. If we have multiple ways to do the reduction, they must be equivalent. More specifically, projected onto the unique block, the reduction operation is unique up to phase (if the final canonical form is fixed, not up to gauge). This phase factor is going to play an important role in our study of SPT orders.

Similar reduction procedure applies when a matrix product unitary operator acts on a matrix product state. In particular, suppose $T^{i,i'}$ is a MPUO and $A^i$ represents a MPS which is symmetric under it. Suppose that $T^{i,i'}$ and $A^i$ are both in the canonical form and have only one block. Because $T^{i,i'}$ represents a symmetry of $A^i$
\be
\t A^i= \sum_{i'} T^{i,i'} \otimes A^{i'}
\ee
represent the same matrix product state as $A^i$. Moreover, because $A^i$ is short range correlated and $T^{i,i'}$ does not increase correlation length, $\t A^i$ is still short range correlated and it also contains one block in its canonical form. However, note that $T^{i,i'}$ is a matrix and the inner dimension of $\t A^i$ is in general larger than that of $A^i$. Therefore $\t A^i$ may no longer be in the canonical form. Some reduction procedure needs to be done to bring $\t A^i$ back to the canonical form.

Suppose that $P$ is the projection onto the single block in the canonical form of $\t A^i$. Due to the uniqueness of the canonical form of a MPS, $P$ must be of the same dimension as $A^i$ and $P\t A^i P$ must be equivalent to $A^i$ up an invertible transformation $Q$.\cite{PVW0701,PWS0802} That is
\be
A^i = QP^{\dg} \t A^i PQ^{-1}
\ee
Denote $V_r=PQ^{-1}$ and $V_l=QP$, we get
$A^i = V_l \t A^i V_r$.
Moreover, $V_lV_r=I$, the identity on the inner dimensions of $A^i$. As $Q$ is unique up to phase, $V_l$ and $V_r$ are unique on the single block of $\t A^i$ up to a conjugate phase factor. With slight abuse of notation, we will denote $V_r$ as $V$ and $V_l$ as $V^{\dg}$ and we have
\be
A^i = V^{\dg} \t A^i V
\ee

\end{document}